\include{psfig}
\documentstyle[epsf]{mn}
\newcommand{\ltsima}{$\; \buildrel < \over \sim \;$}
\newcommand{\simlt}{\lower.5ex\hbox{\ltsima}}            
\newcommand{\gtsima}{$\; \buildrel > \over \sim \;$}
\newcommand{\simgt}{\lower.5ex\hbox{\gtsima}}            
\newcommand{\lya}{Ly$\alpha\,$}
\newcommand{\pks}{PKS\,$0528-250\,$}

\title[HST images of a galaxy group at $z=2.81$]
{HST images of a galaxy group at $z=2.81$, and
the sizes of damped Ly$\alpha$ galaxies
\thanks{Based on observations with the NASA/ESA {\it Hubble
Space Telescope}, and on observations collected at the European Southern
Observatory, La Silla, Chile}}

\author[P. M\o ller and S. J. Warren]
{P. M\o ller$^{\,1,2}$ and S. J. Warren$^{\,3}$\\
$^1\,$ Space Telescope Science Institute, 3700 San Martin Drive,
Baltimore, MD 21218, USA \\
$^2\,$ on assignment from the Space Science Department of ESA \\
$^3\,$ Blackett Laboratory, Imperial College of Science Technology and
Medicine, Prince Consort Road, London SW7 2BZ}

\begin{document}

\maketitle

\begin{abstract}

We present HST WFPC2 observations in three bands (F450W=B, F467M and
F814W=I) of a group of three galaxies at $z=2.8$ discovered in a
ground-based narrow-band search for \lya emission near the $z=2.8$
quasar \pks. One of the galaxies is a damped \lya (DLA) absorber and
these observations bear on the relation between the DLA clouds and the
Lyman-break galaxies and the stage in the evolution of galaxies they
represent. We describe a procedure for combining the undersampled WFPC2
images pointed on a sub-pixel grid, which largely recovers the full
sampling of the WFPC2 point spread function (psf). These three galaxies
have similar properties to the Lyman-break galaxies except that they
have strong \lya emission. The three galaxies are detected in all three
bands, with average $m_B\sim26$, $m_I\sim25$. Two of the galaxies are
compact with intrinsic (i.e. after correcting for the effect of the
psf) half-light radii of $\sim0.1$ arcsec ($0.4 h^{-1}$ kpc,
$q_{\circ}=0.5$). The third galaxy comprises two similarly compact
components separated by 0.3 arcsec. The HST images and a new
ground-based \lya image of the field provide evidence that the three
galaxies are more extended in the light of \lya than in the
continuum. Combined with the evidence from the \lya line widths,
previously measured, this suggests that we are measuring the size of
the surface of last scattering of the escaping resonantly-scattered
\lya photons. The measured impact parameters for this DLA galaxy
(1.17 arcsec), for a second confirmed system, and for several
candidates, provide a preliminary estimate of the
cross-section-weighted mean radius of the DLA gas clouds at $z\sim 3$
of $<13 h^{-1}$ kpc, for $q_{\circ}=0.5$. The true value is likely
substantially smaller than this limit as DLA clouds at small impact
parameter are harder to detect. Given the observed sky covering factor
of the absorbers this implies that for $q_{\circ}=0.5$ the space
density of DLA clouds at these redshifts is more than five times the space
density of spiral galaxies locally, with the actual ratio probably
considerably greater. For $q_{\circ}=0.0$ there is no evidence as yet
that DLA clouds are more common than spiral galaxies locally. We
summarise evidence that filamentary structures occur in the
distribution of galaxies at high redshift.

\end{abstract}


\section{Introduction}

By studying typical galaxies at high redshift, here $z>2$, we can
record the origins of normal galaxies such as our own Milky Way.
Schmidt \cite{sc65} was the first to observe a high$-$redshift galaxy
when he obtained a spectrum of 3C9, a radio$-$loud quasar of redshift
$z=2.01$. Normal star$-$forming high$-$redshift galaxies are about 1000
times fainter, and remained undiscovered until the present decade. A
small number of candidates (some of which may be active galaxies) have
been identified with 4$-$metre telescopes, through the detection of
Ly$\alpha$ emission (e.g. Steidel, Sargent, and Dickinson 1991,
Lowenthal et al. 1991, M\o ller and Warren 1993, Pascarelle et
al. 1996, Francis et al. 1995). Another approach has been to employ
deep broad$-$band imaging to identify candidates by the expected
Lyman$-$limit discontinuity in their spectra (Steidel and Hamilton
1992). The brightest Lyman$-$break galaxies have $m_R\sim 24$ and
spectroscopic confirmation has only become feasible with the completion
of the Keck 10$-$metre telescope. The report by Steidel et al. (1996)
on the results of spectroscopic observations of Lyman$-$break
candidates really marked the beginning of the statistical study of
high-redshift galaxies. They were able to confirm redshifts for 15
star$-$forming galaxies in the range $3.0<z<3.5$. They discovered that
Lyman$-$break galaxies generally show weak Ly$\alpha$ emission, but
that the rest$-$frame ultra$-$violet spectra may be recognised by the
presence of absorption lines characteristic of the spectra of nearby
star$-$forming galaxies. Madau et al. (1996) applied the same
methodology to the Hubble Deep Field data to extend the results to
fainter magnitudes, and their analysis currently provides the best
lower limits to the integrated star formation rate in galaxies at $z>2$. 

Complementary to the searches for starlight from high$-$redshift
galaxies have been the analyses of absorption lines in the spectra of
high$-$redshift quasars. These studies have provided measurements of
the mass density of neutral hydrogen in the universe (e.g. Wolfe et
al. 1986, Lanzetta et al. 1991), and the metallicity of the gas
(e.g. Pettini et al 1994), and how these quantities have changed with
redshift. The analysis of Pei and Fall (1995) of the absorption--line
data reconstructs the global history of star formation, gas
consumption, and chemical enrichment, accounting in a self--consistent
way for the effects of the progressive extinction of the background
quasars due to dust as star formation proceeds.  The advantage of the
absorption--line approach to the history of star formation is that it
is global in nature as all the neutral gas at any redshift is directly
observed. With the deep imaging studies it is necessary to extrapolate
below the survey flux limit to measure the total star formation rate.

The weakness of the global approach however is that it tells us nothing about
the sizes and morphologies of the galaxies in which the star formation
occurs. Deep imaging of DLA absorbers provides the missing
information. In other words if we imaged the absorbers we could combine
the data of faint galaxy surveys (luminosities, sizes, shapes) with the
information on the gas obtained from the spectroscopic studies (column
densities, chemistry), allowing a more detailed comparison with
theories of galaxy formation.

Unlike for MgII absorbers (e.g. Bergeron and Boiss\'e 1991, Steidel,
Dickinson, and Persson 1994a) little progress has been made in such a
programme of imaging of DLA systems. There are only two published
unambiguous detections, discussed below, as well as a small number of
candidate counterparts of DLA absorbers (at both low redshift $-$
Steidel et al. 1994b, Le Brun et al. 1997 $-$ and high-redshift $-$
Arag\'{o}n-Salamanca et al. 1996). In this paper we report on the
results of 30 orbits of imaging observations with the {\it
Hubble Space Telescope} (HST) of a group of three galaxies at
$z=2.81$, named S1, S2, S3, one of which (S1) is a damped \lya
absorber. The three galaxies lie in the field of the quasar \pks, and
were detected in the light of \lya emission by M\o ller and Warren
(1993, hereafter Paper I). A detailed discussion of their nature was
presented by Warren and M\o ller (1996, hereafter Paper II \footnote{We
note here that recent spectroscopic observations by Ge et al. (1997) confirm
our conclusion that the \lya emission from the DLA absorber is due to
star formation rather than photoionisation by the quasar.}). The HST
images provide measures of their sizes, magnitudes, and colours.  The
other high$-$redshift damped system that has been successfully imaged
is the absorber at $z=3.150$, towards the quasar $2231+131$, observed
by Djorgovski et al. \cite{dj96}. In this paper we intercompare the
measured properties of these two absorbers, the two companion
galaxies S2 and S3, and the population of Lyman$-$break galaxies, to draw
conclusions about the space density of DLA absorbers, their
structure, and the relation between the DLA absorbers and the
Lyman-break galaxies.

The layout of the rest of the paper is as follows. The HST
observations of the field towards \pks are described in Section 2, as
well as new ground$-$based narrow$-$band observations of the
field. The three galaxies are very small, and the HST images were
dithered with a half-integer pixel step in order to improve the image
sampling. The algorithm for combining the images is outlined in this
section. In Section 3 we present the reduced HST images. In Section 4
we provide the results of aperture photometry and profile fitting of
the HST and ground-based images. Finally, in Section 5 we discuss
these results and their implications for our understanding of the
nature of DLA absorbers and Lyman-break galaxies.

\section{Observations}

\subsection{HST observations}

The HST observations were made in 1995 February and March, using the
Wide Field Camera of WFPC2. The journal of observations appears in
Table 1. We used three different filters: six orbits each of
observations with the WFPC2 standard broad B (F450W) and I (F814W)
filters, and 18 orbits with the F467M medium-passband filter. The
last filter, hereafter M, has a width of $215 {\rm \AA}$ (FWHM), and
is the narrowest WFPC2 filter that contains the wavelength of
redshifted Ly$\alpha$ of the targets, $4630{\rm\AA}$.

\begin{table}
\begin{flushleft}
\caption[]{Journal of HST observations}
\begin{tabular}{lccc}
\hline\noalign{\smallskip}
\multicolumn{1}{c}{Date}& Filter& Band &Exposure \\
    &       &   & (sec)    \\
\noalign{\smallskip}
\hline\noalign{\smallskip}
1995 Feb 2& F467M & M & 12900 \\
1995 Mar 5& F467M & M & 12900 \\
1995 Mar 8& F814W & I & 12900 \\
1995 Mar 8& F450W & B & 12900 \\
1995 Mar 9& F467M & M & 12900 \\
\noalign{\smallskip}
\hline
\end{tabular}
\end{flushleft}
\end{table}

The 30 orbits were scheduled as five visits, each of six orbits. For
each visit we observed through only one filter. In each case we
obtained a 1900 sec exposure in the first orbit, and an uninterrupted
2200 sec exposure in each of the following 5 orbits. The six exposures
of each visit were grouped in three pairs. The two exposures of each
pair were obtained with exactly the same pointing. Relative to the
first, the second and third pairs were shifted, respectively, by 0.45
arcsec in $x$, and by 0.45 arcsec in both $x$ and $y$. The step of 0.45
arcsec is exactly 4.5 WFC pixels, and this strategy was chosen with a
view to improving the rejection of cosmic rays, and to allow an
increase in the sampling, since the WFPC2 point spread function is
undersampled by the 0.1 arcsec WFC pixels. The latter point is
important because we need good sampling for a morphological study, but
our objects are too faint to be observed with the Planetary Camera.

For the basic reduction of the data we obtained, from the STScI data
archive, the best biases, darks and flats appropriate for the observing
dates and applied these in standard ways. We also obtained the relevant
hot pixel lists, and created a bad pixel mask and a hot pixel mask for
each image.

To combine the frames for each filter we used the weighting scheme and
the sigma clipping algorithm described in Paper I, which is optimal for
faint sources.  We have extended the algorithm to allow for half
integer shifts between input frames. A short description of the
procedure follows. A detailed discussion will be provided elsewhere
(M{\o}ller, in preparation).

The sub$-$pixel combination is undertaken in two stages, of which the
second is iterative.  The first stage is the combination of the
frames using integer pixel shifts, with the rejection of cosmic rays by
sigma clipping, and the masking of bad pixels. In this combined image
each pixel is then divided two by two into sub$-$pixels, providing the
first estimate of the final sub$-$pixelised image. The pixels in the
original data frames are then similarly divided, and each frame is
registered with the combined image, to the nearest sub$-$pixel. The
sub-pixel registration allows improved rejection of cosmic rays in the
subsequent processing, notably where there are strong gradients in the
data e.g. in images of stars.

The first combined image provides a prescription for each original data
pixel as to the distribution of the detected incident photons within
the pixel. In each frame, in each pixel, the counts are then
distributed to the four corresponding sub$-$pixels according to this
prescription. A second combined frame is then constructed, again after
appropriate rejection of bad pixels and cosmic rays. This new frame
provides an improved prescription for the redistribution of counts
within pixels, and the final frame is constructed in this manner
through iteration. The iteration is halted as soon as the mean $\chi^2$
per pixel has reached a stable minimum, typically after six to twelve
iterations. Along with the combined image, the corresponding variance
image is created.

The measured profiles of objects in the final frame will be similar to
the profiles that would have been obtained for the same instrument but
with pixels of side half the size. Thus the algorithm improves the
sampling by a factor of two, but does not attempt to deconvolve the
resolution set by the optics. It does, however, remove the extra
``box--car'' smoothing of the image caused by the finite pixel size.
The algorithm is strictly locally and globally count conserving,
and the observed number of counts recorded in a whole detector pixel
is never changed (as opposed to deconvolution algorithms attempting to
correct for the effect of the psf).

Most cameras used in optical astronomy have pixels which are small
compared to the resolution FWHM of the images, and hence their
intrinsic  pixel smoothing is insignificant. For WFPC2 the pixels
are of the same size as the resolution, so the pixel smoothing adds
significantly to the final resolution if it is not corrected for. The
correction algorithm outlined above therefore, for WFPC2, has the
advantage of improving both sampling and resolution, but it does not
suffer the disadvantage of mixing information between unrelated
detector pixels
as does psf--deconvolution algorithms. The main difference between our
algorithm and that used by the HDF group (Hook and Fruchter, 1997) is
that the HDF group decided not to correct for the pixel smoothing
because the correction introduces a small correlation of noise on the
scale of an original pixel. The HDF images therefore have somewhat
lower resolution but statistically independent sub--pixels.

\subsection{NTT observations}

We obtained a new deep ground-based \lya image of the field using the
SUSI instrument on the ESO New Technology Telescope (NTT), over two
runs in December 1993 and December 1994. The journal of the
observations appears in Table 2. The narrow-band filter used for the
NTT observations is the same filter, hereafter N, as was used for the
ESO 3.6m observations reported in Paper I, and has a central
wavelength $\lambda=4633{\rm\AA}$, FWHM $23{\rm \AA}$.

Our original ESO 3.6m narrow-band image was badly affected by filter
ghosts, and by the poor seeing, average 1.7 arcsec FWHM. The purpose of
the SUSI image was to obtain a deeper ghost-free image (as the filter
is not inclined in SUSI) with better seeing. The observations were
broken into 1 hour exposures obtained with small shifts in
pointing. The data reduction and the combination of frames followed the
procedures described in Paper I. The images in the final narrow-band
frame have FWHM 0.96 arcsec, with a sampling of 0.26 arcsec per
pixel. The source S1 lies at an angular separation of only $\sim1.2$ arcsec
from the line of sight to the quasar. Due to the relatively good
seeing the faint image of the quasar (from the small amount of quasar
light that leaks through the filter wings) and the image of S1 are
separate. There are several unsaturated images of bright stars in the
SUSI frames, so that using DAOPHOT we were able to subtract the image
of the quasar.

\begin{table}
\begin{flushleft}
\caption[]{Journal of NTT/SUSI observations}
\begin{tabular}{lcc}
\hline\noalign{\smallskip}
\multicolumn{1}{c}{Date}& Filter& Exposure \\
    &       & (sec)    \\
\noalign{\smallskip}
\hline\noalign{\smallskip}
1993 Dec 10/11& N & 15700 \\
1994 Nov 30/Dec 1& N & 18000 \\
1994 Dec 1/Dec 2& N & 18500 \\
\noalign{\smallskip}
\hline
\end{tabular}
\end{flushleft}
\end{table}

\section{HST images}

In Figure 1 we show a small section of the field around the quasar,
that includes the three galaxies S1, S2, S3. The figure is a weighted
sum of the final HST frames for the three filters.  The three galaxies
are clearly visible in this figure, and are also each detected in the
individual summed frames for each filter. The galaxy S1 lies at an
angular separation of only $1.17 \pm 0.02$ arcsec from the quasar, and
is only visible after subtraction of the quasar image. In Figure 1 we
have inserted a small panel from the image with the quasar subtracted,
to reveal S1. The lower panel of Figure 1 is the same image lightly
smoothed to enhance the contrast of faint objects.

\begin{figure*}
\vspace{12.0cm}
\includegraphics{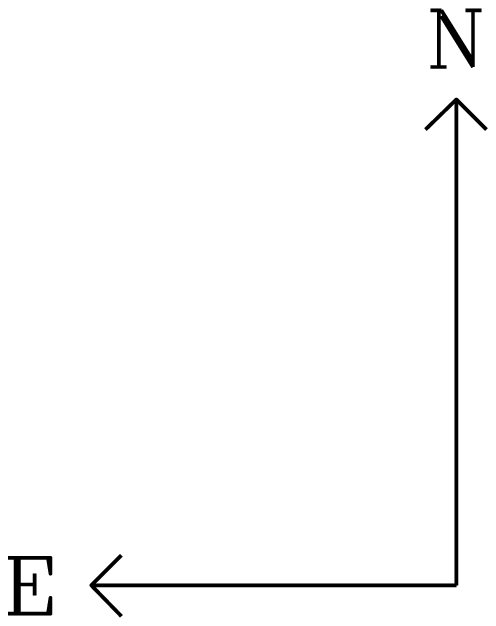}
\includegraphics{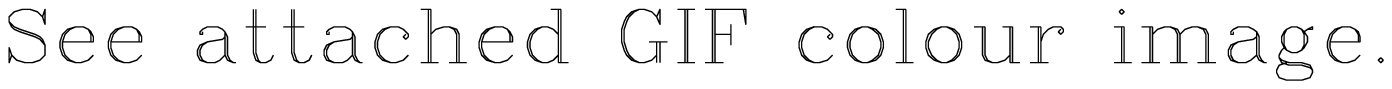}
\caption[ ]{HST Wide Field Camera 2 sub--pixelised image of a region 
$43{\rm\, arcsec}\times 17{\rm\, arcsec}$ around the quasar \pks (marked
Q), showing the three galaxies S1, S2, S3. The size of each sub--pixel
is 0.05 arcsec. The upper panel is a weighted average of the data for
the three filters, from all 30 orbits. The lower panel is the same
image convolved with a Gaussian profile of $\sigma=1.3$ sub--pixel. An
inset panel shows a small region to the S of the quasar after
subtraction of the quasar image, revealing the image of S1. The scale
is shown by a horizontal bar of length 3 arcsec, the orientation is
given by the N/E arrows to the right.}
\end{figure*}

For the psf subtraction we used TinyTIM v4.1 model psfs. Model psfs
were made for each filter, and were constructed on the 0.05 arcsec
sub--pixel scale. Images of S1 in all three bands, before and after psf
subtraction, are shown in Figure 2. Construction of
sub-pixelised images and psfs provides a substantial improvement
in the psf subtraction over our preliminary integral pixel shift
reductions shown in M{\o}ller and Warren (1996).

A comparison between images obtained in different bands is provided in
Figure 3, which shows close--up images of all three galaxies. The
sources S1 and S2 are both compact, without any obvious substructure,
while the source S3 is composed principally of two compact components,
separated by 0.3 arcsec. A possible third component is just visible,
but contains only $4\%$ of the light, and has not been included in the
profile fitting described in the next section.

At this point it is worth commenting on the low signal-to-noise ratio
(S/N) apparent in Figure 3 of the M-band images of sources S2 and S3
compared with the B-band and I-band images.  The \lya emission line
acounts for a large fraction of the M-band flux for the three
sources. As detailed in Section 5 we find evidence that the \lya
emission is more extended than the continuum emission for the three
sources, which would explain the low S/N of the M-band images.

\begin{figure}
\vspace{6.2cm}
\includegraphics{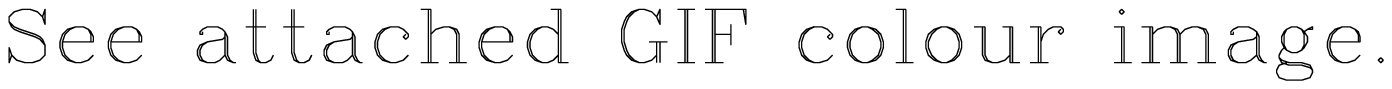}
\caption[ ]{HST Wide Field Camera sub-pixelised images showing the
quasar and the damped \lya galaxy S1. Each box is 3.5 arcsec on a side,
and the size of each sub-pixel is 0.05 arcsec. Top row combined images
for, from left to right, the B, M, and I filters. Bottom row, the same
images after subtraction of the quasar psf, to disclose the image of
S1.}
\end{figure}

\begin{figure*}
\vspace{9.5cm}
\includegraphics{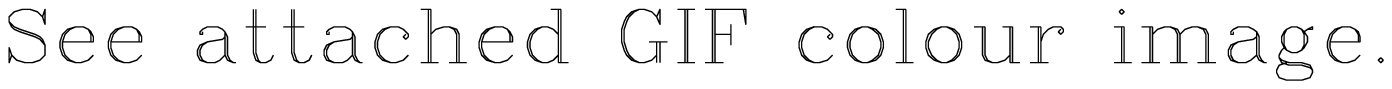}
\caption[ ]{
This figure presents a comparison of images of S1 (top row), S2
(middle row) and S3 (bottom row) in all HST bands. From left to right
the first three columns show the B, M, and I filter images. For the
source S1 the quasar has been subtracted, and the low S/N F467M images
have been lightly smoothed. In column 4 we show a weighted sum of the
images from all three filters, in column 5 our best model fit to the I
band image, and in column 6 the residuals from this fit. All images in
this figure are 2.0 by 2.0 arcsec with orientation as in Figure 1.}
\end{figure*}

\section{Photometry}

\subsection{Aperture photometry}

\begin{table*}
\begin{flushleft}
\caption[]{HST and NTT photometry of S1, S2 and S3}
\begin{tabular}{lcrrrcrr}
\hline\noalign{\smallskip}
\multicolumn{2}{c}{ }  &\multicolumn{3}{c}{Small (isophotal) aperture} &
\multicolumn{3}{c}{Large (circular) aperture} \\
Filter &      & \multicolumn{1}{c}{S1} & \multicolumn{1}{c}{S2} & 
\multicolumn{1}{c}{S3} & \multicolumn{1}{c}{S1$^a$} & 
\multicolumn{1}{c}{S2} & \multicolumn{1}{c}{S3} \\
\noalign{\smallskip}
\hline\noalign{\smallskip}
HST F450W & $m_{\scriptscriptstyle{B}}$      
         & $25.90^{+0.25}_{-0.17}\:\:$ & $26.70^{+0.09}_{-0.08}\:\:$ &
           $25.63^{+0.05}_{-0.05}\:\:$ & $25.45\:\:$ &
           $26.28^{+0.14}_{-0.13}\:\:$ & $25.46^{+0.08}_{-0.08}\:\:$ \\
HST F467M & $m_{\scriptscriptstyle{M}}$    
         & $25.56^{+0.14}_{-0.11}\:\:$ & $26.62^{+0.24}_{-0.20}\:\:$ &
           $25.43^{+0.13}_{-0.12}\:\:$ & $25.18\:\:$ &
           $25.76^{+0.32}_{-0.24}\:\:$ & $25.16^{+0.19}_{-0.16}\:\:$ \\
HST F814W & $m_{\scriptscriptstyle{I}}\:$    
         & $24.96^{+0.13}_{-0.17}\:\:$ & $25.96^{+0.08}_{-0.07}\:\:$ &
           $24.94^{+0.05}_{-0.05}\:\:$ & $24.85\:\:$ &
           $25.66^{+0.16}_{-0.14}\:\:$ & $24.88^{+0.08}_{-0.07}\:\:$ \\
      & $m_{\scriptscriptstyle{B}}$$-$$m_{\scriptscriptstyle{I}}$ 
         & $0.94^{+0.30}_{-0.21}\:\:$  &$0.74^{+0.11}_{-0.11}\:\:$&
           $0.69^{+0.07}_{-0.07}\:\:$  & $0.60$  &
           $0.62^{+0.20}_{-0.21}\:\:$  &$0.58^{+0.11}_{-0.11}\:\:$   \\
      & $m_{\scriptscriptstyle{B}}$$-$$m_{\scriptscriptstyle{M}}$ 
         & $0.34^{+0.27}_{-0.21}\:\:$  &$0.08^{+0.22}_{-0.25}\:\:$&
           $0.20^{+0.13}_{-0.14}\:\:$  & $0.27$  &
           $0.52^{+0.28}_{-0.35}\:\:$  &$0.30^{+0.18}_{-0.21}\:\:$   \\
\noalign{\smallskip}
\hline\noalign{\smallskip}
NTT N   & $m_{\scriptscriptstyle{N}}\:$
               & & & &
               $\:\:\:\:\:\:\:23.07^{+0.10}_{-0.10}$ &
	       $23.48^{+0.16}_{-0.16}\:\:$ &
               $23.17^{+0.11}_{-0.11}\:\:$ \\
      & $m_{\scriptscriptstyle{B}}$$-$$m_{\scriptscriptstyle{N}}$ 
               & & & &
               $2.38\:\:$ &
	       $2.80^{+0.21}_{-0.21}\:\:$ &
               $2.29^{+0.14}_{-0.14}\:\:$ \\
\noalign{\smallskip}
\hline
${\rm EW_{rest}}$& & & & & $\:\:\:\:51$ & $88^{+29}_{-21}\:\:\:\:\:\,$ & 
$46^{+9}_{-7}\:\:\:\:\:\:\,$ \\
Expected &$m_{\scriptscriptstyle{B}}$$-$$m_{\scriptscriptstyle{M}}$ &
& & & 0.56 & $0.77^{+0.12}_{-0.11}\:\:$ & $0.52^{+0.06}_{-0.05}\:\:$ \\
\noalign{\smallskip}
\hline
\end{tabular}

$^a$ B, M, I values measured from the model galaxy profiles
\end{flushleft}
\end{table*}

\subsubsection{Aperture photometry of HST images}

We have measured magnitudes for the three galaxies using both a small
(isophotal) aperture, to obtain accurate colours, and a large
(circular) aperture, to measure total fluxes. The results are provided
in Table 3, as well as the colours $m_B-m_I$, $m_B-m_M$, for each
source, for both apertures. The magnitudes are on the HST system, with
zero points taken from Holtzman et al. \cite{ho95} (their Table 9).

For the small$-$aperture measurements we selected isophotes that
defined areas of 0.13, 0.13, and 0.31 square arcsec for the galaxies
S1, S2, S3, respectively. For each object the identical aperture was
applied in each passband. The quoted photometric errors account for
photon noise, read-out-noise, and the error associated with the
uncertainty in the determination of the local sky level. 

The large aperture used was a circle of diameter one arcsec. The zero
points of the HST photometric system are defined such that for point
sources aperture magnitudes in a circular aperture of this size
provide total magnitudes. Therefore our large-aperture magnitudes will
be total magnitudes if the sources are compact. This is certainly the
case for the continuum emission, as demonstrated by the profile fits
described in the next subsection. The measurement of the flux within
the large aperture is difficult for S1. This is because the aperture
includes regions where the subtraction of the quasar profile is very
uncertain. Instead, for this source we applied the aperture to the
galaxy-profile models described in the next subsection, but have not
attempted to quantify the photometric errors. In computing the
best$-$fit models the bad regions were ignored. Therefore, for S1, the
models ought to provide a more reliable estimate of the flux within
the larger aperture.

\subsubsection{Aperture photometry of NTT image}

For the NTT image we used a circular aperture of diameter 2.6 arcsec. We
measured the aperture correction between this diameter and a very
large diameter for bright stars in the frame, and applied the
correction to the measured magnitudes of our sources. Again, then, if
the sources are compact this procedure provides total magnitudes. The
photometry for the NTT observations is on the AB system.

Also provided in Table 3 are the values of the rest-frame equivalent
width (EW) of the \lya emission line for the three sources. These were
computed by creating power-law spectra $f_{\nu}\propto\nu^0$, with
emission lines of different EW, and applying absorption at wavelengths
shortward of the \lya line centre as approporiate for $z=2.81$, using
the models of M\o ller and Warren (1991). The line EW was adjusted to
reproduce the measured large-aperture $m_B-m_N$ colours.  The value
quoted in the table is the model unabsorbed EW. We used the same model
spectra to compute the expected value of the colour $m_B-m_M$, quoted
in the last row of Table 3. These predicted colours are used in
Section 5.2, in considering the evidence that the \lya emission
is more extended than the continuum emission.

\subsection{Galaxy profile fits}

\begin{table}
\begin{flushleft}
\caption[]{Galaxy profile fits}
\begin{tabular}{lccccc}
\hline\noalign{\smallskip}
\noalign{\smallskip}
 object& band
&$r_{\scriptscriptstyle{0.5}}$&$r_{\scriptscriptstyle{0.5}}$
&$r_{\scriptscriptstyle{0.5}}$&$r_{\scriptscriptstyle{0.5}}$   \\
      &   &de Vauc. &exp.     &average&average$^a$\\
      &   & arcsec  & arcsec  &  arcsec  &kpc \\
\hline\noalign{\smallskip}
 S1   & B &  0.21   &  0.10   &          &             \\
 S1   & M &  0.20   &  0.09   &    \multicolumn{1}{r}{$0.13^b$}  & 
\multicolumn{1}{r}{$0.48h^{-1}$} \\
  S1   & I &  0.13   &  0.08   &\multicolumn{1}{r}{$\pm0.06\:\:$} & 
\multicolumn{1}{r}{$\pm0.21h^{-1}$}   \\
\hline\noalign{\smallskip}
 S2   & B &  0.10   &  0.06   &\multicolumn{1}{r}{$0.08\:\:$}&
\multicolumn{1}{r}{$0.29h^{-1}$}\\
 S2   & I &  0.08   &  0.07   &\multicolumn{1}{r}{$\pm0.01\:\:$} & 
\multicolumn{1}{r}{$\pm0.05h^{-1}$}   \\ 
\hline\noalign{\smallskip}
 S3-a & B &  0.07   &  0.06   &\multicolumn{1}{r}{$0.06\:\:$}&
\multicolumn{1}{r}{$0.23h^{-1}$}\\
 S3-a & I &  0.07   &  0.06   &\multicolumn{1}{r}{$\pm0.01\:\:$} &
\multicolumn{1}{r}{$\pm0.03h^{-1}$}   \\
\hline\noalign{\smallskip}
 S3-b & B &  0.13   &  0.09   &\multicolumn{1}{r}{$0.11\:\:$}&
\multicolumn{1}{r}{$0.42h^{-1}$}\\
 S3-b & I &  0.13   &  0.10   &\multicolumn{1}{r}{$\pm0.02\:\:$} &  
\multicolumn{1}{r}{$\pm0.08h^{-1}$}   \\
\noalign{\smallskip}
\hline
\end{tabular}
\end{flushleft}

$^ah=H_{\circ}/100$, $q_{\circ}=0.5$ \\
$^b$ average of values for B and I bands only \\
\end{table}

\subsubsection{Profile fits for HST images}

We have measured the half-light radii $r_{\scriptscriptstyle{0.5}}$ of
the galaxies S1 (after subtraction of the quasar image) and S2, and of
the two components of the galaxy S3 by fitting de Vaucouleurs and
exponential profile functions. The results are provided in Table 4.
The two components of S3 are separated by $0.30\pm0.01$ arcsec. The
S/N of the M-band images of S2 and of S3 were too low to provide
meaningful results. There follows a description of the fitting procedure.

Firstly any residual sky counts (or large-scale psf residuals for S1)
in the frames were subtracted by means of a large-scale running modal
filter, where the mode is estimated by fitting a Gaussian to the peak
in the histogram of the counts in the sliding box. A box of side 51
sub-pixels (2.55 arcsec) was used for S2 and S3. A smaller box of side
41 sub-pixels was used for S1, ignoring pixels close to the quasar
centroid. For the profile fits themselves boxes of side 21 sub-pixels
were used for S2 and S3, and of side 15 sub-pixels for S1. The results
were fairly insensitive to the choices of box size. The model profiles
are characterised by six parameters: $x, y$, ellipticity, position
angle, $r_{\scriptscriptstyle{0.5}}$, and surface brightness at
$r_{\scriptscriptstyle{0.5}}$. A mask was applied to exclude certain
pixels from the fit, as required; for example the faint third source
in fitting S3, and in fitting S1 the region over which the quasar
image is bright (i.e. where the Poisson noise is high, and where large
residuals from the psf subtraction remain).  Trial model galaxy light
profiles were convolved with the normalised sub-pixelised light
profile of a nearby star, and the minimum $\chi^2$ fit was found for
the six free parameters.  Examples of the best$-$fit galaxy models are
shown in Figure 3, as well as the images after subtraction of the
models.

Referring to Table 4 it can be seen that for each galaxy, for a
particular profile function, there is good agreement in the measured
half-light radii for the different filters, indicating that the point
spread function is adequately sampled in the sub-pixelised images. The
goodness of fit of both profile functions is in all cases similar, so
there is no indication that either the de Vaucouleurs or exponential
function is preferred. In all cases a larger value of
$r_{\scriptscriptstyle{0.5}}$ is measured for the de Vaucouleurs
function, and the discrepancy is larger than the scatter for the same
function for different filters. This is particularly noticeable for
galaxy S1 and it is possible that this is caused by residuals left from
imperfect subtraction of the quasar image.  The purpose of fitting de
Vaucouleurs and exponential functions is to be able to correct for the
psf. The profile fitting indicates that the dominant uncertainty in the
measurement of the intrinsic half-light radius is the choice of profile rather
than Poisson noise. Therefore the range of half-light radii quoted in
Table 4 for the different profile functions provides a measure of how accurately the
half-light radius is determined. For this reason we have chosen to average
all the measurements as the best estimate of the half-light radius, and
to quote the scatter amongst the measurements for each object as a
measure of the uncertainty, as listed in column 5 of the table.

The psf has a significant effect on the measured half-light radii for
the smallest objects. For example the half-light radius for S2 measured
from the images, without correction for the psf, is 0.15 arcsec, which
is about double the intrinsic value.

\subsubsection{Profile fits for NTT image}

The N-band ground-based images are of higher S/N but lower spatial
resolution than the HST M-band data. Because the continuum contributes
only a small fraction of the light to the N-band images the N-band data
essentially provide the \lya profiles of the three sources, and we
attempted to measure the profiles following the fitting procedure
described above. For S2 and S3 the results are not particularly useful
as evidence of whether the \lya emission is more extended than the
continuum emission. For both sources the measured intrinsic sizes are
consistent with the continuum sizes measured from the HST images, but
also consistent with the hypothesis that they are more extended. The
N-band image of S1 on the other hand is clearly extended. We measured
half-light radii of $0.51^{+0.08}_{-0.08}$ arcsec and
$0.66^{+0.25}_{-0.19}$ arcsec for the exponential and de Vaucouleurs
profiles respectively. The quoted errors are computed as follows: the
value of the radius is varied in steps away from the best--fit value,
and for each fixed value of the radius a new minimum $\chi^2$ fit is
computed by allowing the other parameters to vary. The radius at which
the $\chi^2$ is 1.0 greater than the best--fit $\chi^2$ is
identified. The change in radius relative to the best-fit value is the
$1\sigma$ error. For each object the results of the profile fitting
demonstrate that the 2.6 arcsec diameter aperture used for the
photometry includes most of the light from the source, so we will treat
the measurements as total magnitudes.

\section{Discussion: DLA absorbers and Lyman-break galaxies}

\subsection{Continuum emission}

The three galaxies S1, S2, S3 are similar in their properties to
members of the recently detected population of Lyman-break galaxies,
except that they have strong \lya emission, with mean restframe ${\rm
EW=60\AA}$. For example the B and I magnitudes for our three sources
$m_B\sim 26$, $m_I\sim 25$, are within the range of magnitudes of the
Lyman-break galaxies of redshift $z\sim3$ observed by Steidel et
al. (1996) and Lowenthal et al. (1997). The average half-light radius
of the four objects listed in Table 4 is 0.1 arcsec. This is smaller
than the value quoted for the Lyman-break galaxies (Giavalisco et
al. 1996, Lowenthal et al. 1997), by a factor of two to three. However
we have broken S3 into two sub components and measured the radii for
each subcomponent. In addition the radii quoted in Table 4 have been
corrected for the effect of the psf.

Young star-forming galaxies should have approximately constant flux
per unit frequency over the restframe ultraviolet and optical regions
of the spectrum. The expected colour of a source at $z=2.8$ with a
power-law continuum varying as $f_{\nu}\propto\nu^0$, after accounting
for typical absorption in the \lya forest, is $m_B-m_I=0.66$. As seen
in Table 3 the $m_B-m_I$ colours for the three sources S1, S2, S3 are
consistent with this value, for both the small and large
apertures. 
 
The fact that S1 has properties similar to the Lyman-break galaxies
suggests that DLA absorbers and Lyman-break galaxies may commonly be
associated with each other. Lowenthal et al (1997) have also made a
connection between the two populations by suggesting that the broad
\lya absorption line seen in several cases in the spectra 
of Lyman-break galaxies is a damped line from neutral hydrogen in the
galaxy. However the absorption lines seen in their average spectrum and
in the relatively high S/N spectrum presented by Ebbels et al. (1996)
are not optically thick in the line centre. The absorption--line
profiles are difficult to interpret because the background source is
extended.  It is even possible that in some cases the absorption line
is stellar absorption in the spectra of late B stars and that the
galaxies are observed in a post-burst phase, rather than actively
forming stars (see Valls-Gabaud 1993) i.e. for some cases this may be
the explanation for the weakness of the \lya emission line in the
Lyman-break galaxies, as opposed to dust.  High-redshift galaxies with
strong \lya emission, like the three sources near \pks, may therefore
be younger than many of the Lyman-break galaxies.

\subsection{Extended \lya emission}

In this subsection we summarise the evidence that the \lya emission
from the three sources is more extended than the continuum emission.

The evidence for S2 and S3 comes from an examination of the $m_B-m_M$
and $m_B-m_N$ colours, summarised in Table 3. The bottom row in Table
3 provides the predicted $m_B-m_M$ colours for each source computed,
as detailed in Section 4.1.2, from the B and N total magnitudes. The
small aperture $m_B-m_M$ colours are smaller than the predicted
values, as would be the case if the \lya emission is more extended
than the continuum. For S2 the difference is significant at
$2.8\sigma$, and for S3 at $2.3\sigma$. Notice also that the large
aperture $m_B-m_M$ colours are larger than the small aperture colours,
which is consistent with the hypothesis of extended \lya emission. A
broad absorption line in the M passband could only explain part of the
discrepancy between the measured and predicted colours. For example
for S2, simply removing $50{\rm\AA}$ of the continuum flux within the
M band changes the $m_B-m_M$ colour by only 0.1 mag.

The total B magnitude for S1 (i.e. the large-aperture measurement from
the HST image, Table 4) is not reliable because of the uncertainty
associated with the psf subtraction. This prevents us from usefully
making the same comparison between predicted and measured $m_B-m_M$
colours for S1. However there is an indication that S1 may also be
more extended in \lya. The measurements of the half-light radius of S1
from the ground-based \lya image yielded values of
$0.51^{+0.08}_{-0.08}$ arcsec for the exponential profile and
$0.66^{+0.25}_{-0.19}$ arcsec for the de Vaucouleurs profile (Section
4.2.2). These values are significantly larger than the half light
radius of the region of continuum emission of $0.13\pm0.06$ arcsec
(Table 4), measured from the HST image. Such a diffuse contribution of
\lya to the M-band flux would be hard to detect, and could have been
partially subtracted in attempting to remove large-scale residuals
from the psf subtraction (Section 4.2.1). However we cannot rule out
the possibility that there is also a similar diffuse contribution to
the continuum flux.

The argument for extended \lya is made visually in Figure 4. Here we
compare the M-band images after subtraction of the computed
contribution of the continuum (i.e. showing the \lya emission only),
against how the objects would have appeared if the
\lya light profile were the same as the continuum profile.

\begin{figure}
\vspace{8.5cm}
\includegraphics{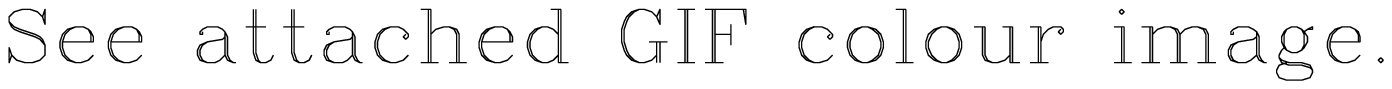}
\caption[ ]{Illustration of the evidence for extended \lya emission
for, from top to bottom, S1, S2, S3. Each panel shows the same 2
arcsec by 2 arcsec region as Figure 3, here smoothed by a 0.15 by 0.15
arcsec boxcar filter to enhance faint features. The left column shows
the summed image from all three filters. The middle column shows the
M-band images with the computed contribution of the continuum
subtracted. This image, then, contains only the \lya line component.
In the right column we show how the \lya-only image would
have appeared if the \lya profile were identical to the continuum
profile.}
\end{figure}

To summarise, there is evidence that the regions of \lya emission from
these three sources are more extended than the regions of continuum
emission. Although the colour and size differences quoted above are
only marginally significant, these results accord with our earlier
conclusion (Paper II) that the cause of the relatively broad \lya
emission lines for S1 and S3 is resonant scattering, as the photons
escape though a high column density of HI. In this picture the \lya
image records the last scattering surface, which will be larger than
the region of star formation. We note in passing that this implies that
the observed strength of \lya absorption and emission in spectra can
depend on the slit width used.

\subsection{The sizes of the DLA gas clouds}

In this section we make a summary of the measured impact parameters
$b$ of DLA absorbers. Using this we estimate the
cross-section-weighted mean radius $\bar{R}_{DLA}$ of the gas clouds
at $z>2$. Combining this information with the measured line density of
DLA absorbers $dn/dz$ we are able to infer the ratio of the comoving
space density of DLA absorbers at $z>2$ to the local space density of
spiral galaxies.

\begin{table*}
\begin{flushleft}
\caption[]{Measured impact parameters $b$ of DLA absorbers}
\begin{tabular}{lcccccc}
\hline\noalign{\smallskip}
Quasar&$z_{abs}$&log(N(HI))&$b$&$b$& confirmed & reference \\
  &  &  &($q_{\circ}=0.0$)&($q_{\circ}=0.5$)& & \\
      &         &cm$^{-2}$ &kpc&kpc& & \\  \hline 
$0454+0393$ & 0.86 & 20.8  & $4.1h^{-1}$& $3.3h^{-1}$ & N & 1  \\
$0302-223$  & 1.01 &$\leq20.0\:\:\:\:\:$& $6.2h^{-1}$& $4.9h^{-1}$ & N & 1  \\
$1331+170$  & 1.78 & 21.2  & $4.7h^{-1}$& $3.1h^{-1}$ & N & 1  \\
$0151+048$A & 1.93 & 20.4  & $7.7h^{-1}$& $5.0h^{-1}$ & Y & 2  \\
$1215+333$  & 2.00 & 21.0  & $8.4h^{-1}$& $5.3h^{-1}$ & N & 3  \\
$0841+129^a$& 2.37 & 21.3  & $8.0h^{-1}$& $4.7h^{-1}$ & N & 3,4  \\
$0841+129^a$& 2.48 & 21.0  & $8.0h^{-1}$& $4.7h^{-1}$ & N & 3,4  \\
$0528-250^b$& 2.81 & 21.3  & $8.1h^{-1}$& $4.5h^{-1}$ & Y & 5  \\
$2231+131$  & 3.15 & 20.0&$15.7h^{-1}\:\,$& $8.2h^{-1}$ & Y & 5  \\
\noalign{\smallskip}
\hline
\end{tabular}
\\
$^a$ the detected galaxy could be the counterpart to one or other of
the two DLA absorbers listed \\
$^b$ impact parameter taken from HST image (this paper) \\
References: 1. Le Brun et al. (1997), 2. Fynbo et al. (1997),
3. Arag\'{o}n-Salamanca et al. (1996), 4. Wolfe et al. (1995),
5. M\o ller and Warren (1993), 6. Djorgovski et al. (1996).
\end{flushleft}
\end{table*}

There have been many searches for optical counterparts of the DLA
absorbers.  In Table 6 we have listed the small number of DLA
absorbers, with redshifts $z>0.8$, for which likely candidate or
confirmed optical counterparts have been detected.  Listed are the
quasar name, the absorber redshift and column density, the measured
impact parameter (i.e. the projected physical separation between the
optical counterpart and the line of sight to the quasar), and whether
or not the counterpart has been confirmed, i.e. the redshift of the
counterpart has been measured to be the same as the redshift of the
absorber.

In Figure 6 we plot impact parameter against column density for the
absorbers listed in Table 6, for $q_{\circ}=0.5$. The detection of the
counterpart to a DLA absorber by broad-band imaging, which requires the
digital subtraction of the quasar image, is more difficult for small
impact parameters where the photon noise from the quasar image is
greater.  Therefore it is very likely that the measured impact
parameters of the few absorbers for which counterparts have been
detected are larger than the average impact parameter for DLA
absorbers. The vertical line in Figure 6 at ${\rm log(N(HI))=20.3}$
marks the conventional lower limit of the column density of DLA
absorbers.  The distribution of points in the figure is consistent with
the expected anticorrelation between impact parameter and column
density. We are interested in the mean impact parameter $\bar{b}_{DLA}$
that would be measured if counterparts were detected for all DLA
absorbers in a large unbiased survey.  From inspection of Figure 6 we
suggest that it is safe to conclude that for $z>2$, $q_{\circ}=0.5$,
$\bar{b}_{DLA}$ is less than $7h^{-1}$ kpc. For $q_{\circ}=0.0$ we
suggest $\bar{b}_{DLA}<13h^{-1}$ kpc. The actual mean impact parameters
are probably substantially smaller than these limits.

We relate the mean impact parameter to the cross-section-weighted mean
radius $\bar{R}_{DLA}$ of the absorbers by
$\bar{b}_{DLA}=\alpha\bar{R}_{DLA}$. The edge of the DLA is the point
at which the column density falls below $N(HI)=2\times10^{20}$
cm$^{-2}$, and we effectively assume that the column density falls off
sharply at larger radii.  For face on disks $\alpha=\frac{2}{3}$, and
for a disk that is nearly edge on $\alpha=\frac{4}{3\pi}$, so we take
$\alpha=0.55$ as an average value for randomly inclined disks. On this
basis we infer the following limits on the cross-section-weighted mean
radius of DLA absorbers at high redshift:
$$\bar{R}_{DLA}<23.6h^{-1} {\rm kpc}\:\:\:\: (z>2, q_{\circ}=0.0) $$
$$\bar{R}_{DLA}<12.7h^{-1} {\rm kpc}\:\:\:\: (z>2, q_{\circ}=0.5). $$

Using these limits and the measured line density of absorbers $dn/dz$
we can compute limits to the ratio of the comoving space density of
DLA absorbers to the the local space density of spiral galaxies. We
follow essentially the methodology employed by Wolfe et al. (1986),
and Lanzetta et al. (1991, hereafter LTW). For spiral galaxies locally
they adopted a galaxy luminosity function $\Phi(\frac{L}{L_*})$ of
Schechter form, with power-law index $s$, a power-law (Holmberg)
relation between radius and luminosity, of index $t$, and a ratio
between gas radius and optical (Holmberg) radius $\xi$, independent of
luminosity. They found that the incidence of DLA absorbers per unit
redshift $dn/dz$ at $z\sim 2.5$ was considerably higher than expected,
by a factor $F\sim 5$, on the basis of no evolution in galaxy cross
section or luminosity function normalisation. We now allow for
evolution by supposing that the space density of DLA absorbers is
higher than the local space density of spirals by a factor
$E_{\Phi}(z)$, and that the gas radii of galaxies are larger at high
redshift by a factor $E_r(z)$. Since $dn/dz$ is proportional to the
product of the space density and the galaxy cross section $\sigma$, we
have that $F=E_{\Phi}(z)E^2_r(z)$. By comparing the expected value of
the cross-section-weighted radius $\bar{R}$ to the measured limits to
$\bar{R}_{DLA}$, we obtain limits to $E_r(z)$. Then from the measured
values of $F$ we determine limits to $E_{\Phi}(z)$, which is our goal.

Under the above assumptions the cross-section-weighted average
radius of DLA absorbers is given by:
$$\bar{R}(z)=\frac{\int_0^{\infty} R(\frac{L}{L_*})\sigma(\frac{L}{L_*})
\Phi(\frac{L}{L_*})d(\frac{L}{L_*})}
{\int_0^{\infty}\sigma(\frac{L}{L_*})\Phi(\frac{L}{L_*})d(\frac{L}{L_*})}$$
$$=\frac{E_r(z)\xi R_*\Gamma(1+3t-s)}{\Gamma(1+2t-s)}$$
where $R_*$ is the optical radius of a local $L_*$ spiral galaxy.

Following LTW we adopt the following values of the parameters:
$t=0.4$, $s=1.25$, $\xi=1.5$, $R_*=11.5h^{-1}$ kpc. This leads to
$\bar{R}(z)=11.0h^{-1}E_r(z)$.  Comparing with the above
measured limits to $\bar{R}_{DLA}$ we obtain the following limits to
the growth factor of galaxy disks:

$$E_r(z)<2.15\:\: (z>2, q_{\circ}=0.0)$$
$$E_r(z)<1.16\:\: (z>2, q_{\circ}=0.5)$$

For their sample D2, of which at least 30 out of the 38 candidate DLA
absorbers have been confirmed, LTW found $F=3.8$, for $q_{\circ}=0.0$,
and $F=7.1$, for $q_{\circ}=0.5$. These results then imply that the
ratio of the comoving space density of DLA absorbers at high redshift
to the local space density of spiral galaxies is given by:
$$E_{\Phi}(z)=\Phi_*(DLA)/\Phi_*(spiral)>0.8\:\: (z>2, q_{\circ}=0.0) $$
$$E_{\Phi}(z)=\Phi_*(DLA)/\Phi_*(spiral)>5\:\: (z>2, q_{\circ}=0.5) $$
Because the actual average impact parameters of DLA absorbers are very
likely substantially smaller than the quoted limits, the actual
comoving space density ratios are probably considerably greater than
the limits quoted above.

\begin{figure}
\vspace{11.5cm}
\includegraphics{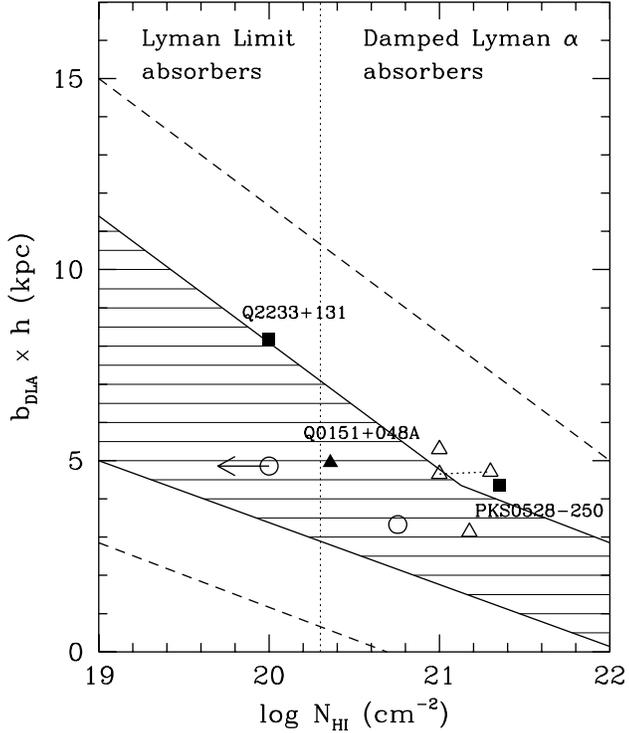}
\caption[ ]{Relation between impact parameter and column density for
all confirmed or candidate counterparts of DLA absorbers of redshift
$z>0.8$ (listed in Table 6). Different symbols correspond to different
redshift ranges: squares $z>2.5$; triangles $2.5\geq z>1.5$; circles
$1.5\geq z>0.8$. Absorbers for which the counterpart has been
confirmed by spectroscopy are shown as filled symbols, and are labeled. 
The two possible DLA absorbers for the quasar $0841+129$ are joined by
a dotted line. The hatched area shows the densely populated region
from the simulation of Katz et al. (1996), and the dashed lines are
the boundaries enclosing all the points in the simulation.}
\end{figure}

In Figure 5 we show also the relation between impact parameter and
column density measured by Katz et al (1996) from a hydrodynamic
simulation of a CDM universe with $q_{\circ}=0.5$. Given the limited
spatial resolution of the simulation, and the fact that star formation
and consequent feedback were not treated, it would be premature to
draw any conclusions about the apparent good agreement between the
results of the simulation and the observations. Nevertheless this plot
and the above calculation underscore the importance of measuring
impact parameters for a large sample of DLA absorbers, a goal we are
pursuing by imaging with the STIS instrument on HST.

\subsection{The structure of the DLA absorbers}

The observations summarised above provide some indication of the
typical structure of a DLA absorber. The following sketch is
suggested. At the centre of the gas cloud, corresponding in projection
to the highest column densities, there may be a region of star
formation, of diameter $\sim 1h^{-1}$ kpc. This central source would
be observed as a Lyman-break galaxy. The gas column density decreases
outwards, as suggested by the anticorrelation between impact parameter
and $N_{HI}$. The size of the region over which the column density is
$>2\times10^{20}$cm$^{-2}$ is several kpc. The region of star formation would
be surrounded by a zone of ionised gas. The \lya photons are
resonantly scattered in escaping to the surface of the surrounding
cloud of neutral gas. The size of the observed region of \lya emission
is larger than the region of star formation, but smaller than the
diameter of the gas cloud. The latter indicates a preferred direction
of escape, implying that the gas resides in a flattened structure. In
reality the gas cloud is likely to be irregular rather than smooth,
and might contain several knots of star formation, surrounded by HII
regions of different sizes, and the \lya photons will escape
preferentially where the column density of neutral gas is lowest,
possibly along a complex network of tunnels.  Merging clouds will be
observed as galaxies with irregular structure.

\subsection{Filamentary structure in the distribution of galaxies at
$z=3$}

We have previously suggested (Paper II) that the approximately linear
arrangement of S1, S2, S3, as well as other known groups at
high-redshift, may correspond to the filamentary arrangements of
galaxies and galaxy sub-units that are found in computer simulations
of the high-redshift universe. Figure 6 summarises the observational
situation, and shows the spatial arrangement of four high-redshift
groups of galaxies discovered in \lya searches (from top to bottom:
Francis et al., 1995, considering the \lya absorber as a galaxy;
Pascarelle et al., 1996; Paper I; Le F\`{e}vre et al., 1996). All
groups have been rotated to a horizontal baseline. Clearly
each of the structures is elongated.

\begin{figure}
\vspace{9.5cm}
\includegraphics{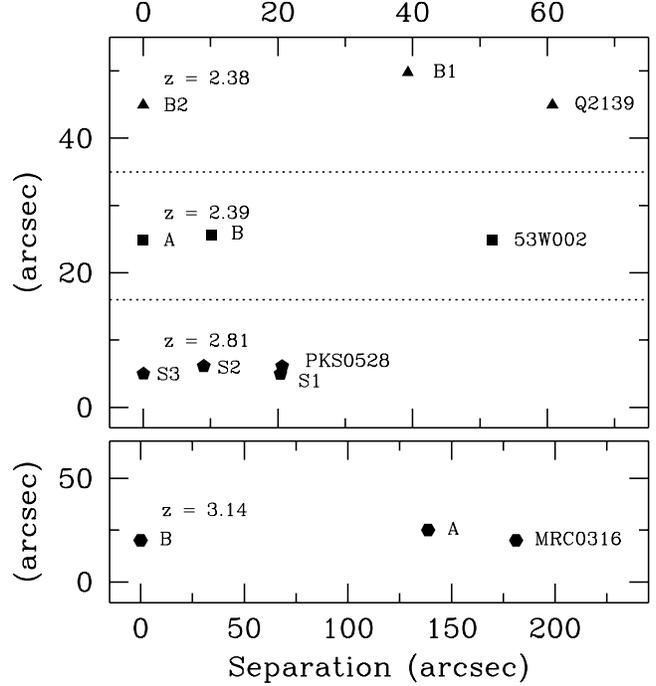}
\caption[ ]{Relative positions of galaxies in four high-redshift groups
discovered in \lya searches. Each group has been rotated so that the
principal axis is horizontal in the figure, and the groups have been
offset from one another vertically. Note the different scale of the
lower plot.}
\end{figure}

To quantify the degree of alignment of each group we have measured the
smallest internal angle $\delta_L$ of the triangle defined by the
three objects in each group (for the PKS 0528-250 field at z=2.81 we
used the position of S1 for the DLA absorber, rather than the position
of the quasar sightline through the absorber). The measured angles
are, from top to bottom in the figure, $\delta_L = 7.0^{\circ},
1.0^{\circ}, 5.8^{\circ}, 2.1^{\circ}$. The average value of the angle
$\delta_L = 4.0^{\circ}$ is certainly very suggestive of
filamentariness in the distribution of galaxies at high redshift. It
is also notable that the sizes of the groups range over an order of
magnitude. In the computer simulations filaments are seen at all
scales (e.g. Evrard et al. 1994). We do not suggest that every
high-redshift group of three or four galaxies discovered will exhibit
a similar degree of alignment, but rather that filamentary networks
akin to those seen in the simulations may become visible as larger
samples of high-redshift galaxies with denser sampling become
available. The phenomenon might provide a useful discriminant for
models of structure formation.

\section*{Acknowledgments}
We thank Johan Fynbo for help with the data reduction, Paul Hewett,
with whom the profile fitting software was developed in collaboration,
and Michael Fall for a critical reading of the paper.
We thank an anonymous referee for severel useful comments which helped
us clarify the paper on several points.


\begin{thebibliography}{}
\bibitem[1996]{ar96} Arag\'{o}n-Salamanca A., Ellis R. S., O'Brien
K. S., 1996, MNRAS 281, 945 
\bibitem[1991]{be91} Bergeron J., Boiss\'e P., 1991, A\&A 243, 344
\bibitem[1996]{dj96} Djorgovski S. G., Pahre M. A., Bechtold J., Elston
   R., 1996, Nature 382, 234
\bibitem[1996]{eb96} Ebbels T. M. D., Le Borgne J.-F., Pell\`{o} R., Ellis
R. S., Kneib J.-P., Smail I., Sanahuja B., 1996, MNRAS 281, L75
\bibitem[1994]{ev94} Evrard A. E., Summers F. J., Davis M. 1994,
ApJ 422, 11
\bibitem[1997]{fy97} Fynbo J., M\o ller P., Warren S. J., 1997, in preparation
\bibitem[1996]{fr96} Francis P. J., Woodgate B. E., Warren S. J.,
   et al., 1995, ApJ 457, 490
\bibitem[1996]{ge97} Ge J, Bechtold J., Walker C., Black J. H., 1997,
ApJ 486, 727
\bibitem[1996]{gi96} Giavalisco M., Steidel C. C., Macchetto F. D.,
   1996, ApJ 470, 189
\bibitem[1995]{ho95} Holtzman J. A., Burrows C. J., Casertano S., et al.,
1995, PASP 107, 1065
\bibitem[1997]{ho97} Hook R. N., Fruchter A. S., 1997, ADASS VI, eds
Hunt G., Payne H. E. (ASP Conf. Ser. 125), p. 147
\bibitem[1996]{ka96} Katz N., Weinberg D. H., Hernquist L.,
   Miralda-Escud\'e J., 1996, ApJ 457, L57
\bibitem[1991]{la91} Lanzetta K. M., Wolfe A. M., Turnshek D. A., Lu
   L., Mc\,Mahon R. G., Hazard C., 1991, ApJS 77, 1
\bibitem[1997]{lb97} Le Brun V., Bergeron J., Boiss\'{e} P., Deharveng
J. M., 1997, A\&A 321, 733
\bibitem[1996]{le96} Le F\`{e}vre O., Deltorn J. M., Crampton D.,
   Dickinson M., 1996, ApJ 471, L11
\bibitem[1991]{lo91} Lowenthal J.D., Hogan C.J., Green R.F., Caulet A.,
   Woodgate B.E., Brown L., Foltz C.B., 1991, ApJ 377, L73
\bibitem[1997]{lo97} Lowenthal J.D., Koo D. C., Guzm\'{a}n R., et al.,
1997, ApJ 481, 673
\bibitem[1996]{ma96} Madau P., Ferguson H. C., Dickinson M. E.,
Giavalisco M., Steidel C. C., Fruchter A., 1996, MNRAS 283, 1388
\bibitem[1991]{mo91} M\o ller P., Warren S. J., 1991, in The Space 
Distribution of Quasars, ed. Crampton D. (ASP Conf. Ser. 21), p. 96
\bibitem[1993]{mo93} M\o ller P., Warren S. J., 1993, A\&A 270, 43
   (Paper I)
\bibitem[1996]{mo96} M\o ller P., Warren S. J., 1996,
   in {\it Cold gas at high redshift}, eds. M.N. Bremer, P.P. van der
   Werf, H.J.A. R\"{o}ttgering and C.L. Carilli, Kluwer Academic
   Publishers, p233
\bibitem[1996]{ps96} Pascarelle S. M., Windhorst R. A., Driver S. P.,
   Ostrander E. J., Keel W. C., 1996, ApJ 456, L21
\bibitem[1995]{pt94} Pettini M., Smith L. J., Hunstead R. W., King
D. L., 1994, ApJ 426, 79
\bibitem[1995]{pe95} Pei Y. C, Fall S. M., 1995, ApJ 454, 69
\bibitem[1965]{sc65} Schmidt M., 1965, ApJ 141, 1295
\bibitem[1991]{st91} Steidel C. C., Sargent W. L. W., Dickinson M.,
   1991, AJ 101, 1187
\bibitem[1992]{st92} Steidel C. C., Hamilton, D., 1992, AJ 104, 941
\bibitem[1994]{st94a} Steidel C. C., Dickinson M., Persson S. E.,
   1994a, ApJ 437, L75
\bibitem[1994]{st94b} Steidel C. C., Pettini M., Dickinson M., Persson
S. E., 1994b, AJ 108, 2046
\bibitem[1996]{st96a} Steidel C. C., Giavalisco M., Pettini M.,
   Dickinson M., Adelberger K. L., 1996, ApJ 462, L17
\bibitem[1996]{va93} Valls-Gabaud D., 1993, ApJ 419, 7
\bibitem[1996]{wa96} Warren S. J., M\o ller P., 1996, A\&A 311, 25
   (Paper II)
\bibitem[1986]{wo86} Wolfe A. M., Turnshek D. A., Smith H. E., Cohen
R. D., 1986, ApJS 61, 249
\bibitem[1995]{wo95} Wolfe A. M., Lanzetta K. M., Foltz C. B., Chaffee
F. H., 1995, ApJ 454, 698
\end{thebibliography}
\end{document}